\documentclass[11pt,a4paper]{article}
\usepackage{jcappub}
\usepackage{graphicx}
\usepackage{dcolumn}
\usepackage{amsfonts}
\usepackage{amssymb}
\usepackage{bm}
\usepackage{hyperref}
\usepackage{url}
\usepackage{subfigure}
\usepackage[sort&compress]{natbib}
\usepackage{txfonts}
\usepackage{color}

\begin{document}
\newcommand{\changeR}[1]{\textcolor{red}{#1}}

\newcommand{\TBB}{{{T_{\rm BB}}}}
\newcommand{\TCMB}{{{T_{\rm CMB}}}}
\newcommand{\Te}{{{T_{\rm e}}}}
\newcommand{\Teq}{{{T^{\rm eq}_{\rm e}}}}
\newcommand{\Ti}{{{T_{\rm i}}}}
\newcommand{\nB}{{{n_{\rm B}}}}
\newcommand{\nHe}{{{n_{\rm ^4He}}}}
\newcommand{\nHet}{{{n_{\rm ^3He}}}}
\newcommand{\nHt}{{{n_{\rm { }^3H}}}}
\newcommand{\nHtw}{{{n_{\rm { }^2H}}}}
\newcommand{\nBes}{{{n_{\rm { }^7Be}}}}
\newcommand{\nLis}{{{n_{\rm { }^7Li}}}}
\newcommand{\nLisi}{{{n_{\rm { }^6Li}}}}
\newcommand{\nS}{{{n_{\rm s}}}}
\newcommand{\Teff}{{{T_{\rm eff}}}}

\newcommand{\id}{{{\rm d}}}
\newcommand{\aR}{{{a_{\rm R}}}}
\newcommand{\bR}{{{b_{\rm R}}}}
\newcommand{\neb}{{{n_{\rm eb}}}}
\newcommand{\neql}{{{n_{\rm eq}}}}
\newcommand{\kB}{{{k_{\rm B}}}}
\newcommand{\EB}{{{E_{\rm B}}}}
\newcommand{\zmin}{{{z_{\rm min}}}}
\newcommand{\zmax}{{{z_{\rm max}}}}
\newcommand{\YBEC}{{{Y_{\rm BEC}}}}
\newcommand{\YSZ}{{{Y_{\rm SZ}}}}
\newcommand{\rhob}{{{\rho_{\rm b}}}}
\newcommand{\Ne}{{{n_{\rm e}}}}
\newcommand{\sigT}{{{\sigma_{\rm T}}}}
\newcommand{\me}{{{m_{\rm e}}}}
\newcommand{\nBB}{{{n_{\rm BB}}}}

\newcommand{\kD}{{{{k_{\rm D}}}}}
\newcommand{\KC}{{{{K_{\rm C}}}}}
\newcommand{\KdC}{{{{K_{\rm dC}}}}}
\newcommand{\Kbr}{{{{K_{\rm br}}}}}
\newcommand{\zdC}{{{{z_{\rm dC}}}}}
\newcommand{\zbr}{{{{z_{\rm br}}}}}
\newcommand{\aC}{{{{a_{\rm C}}}}}
\newcommand{\adC}{{{{a_{\rm dC}}}}}
\newcommand{\abr}{{{{a_{\rm br}}}}}
\newcommand{\gdC}{{{{g_{\rm dC}}}}}
\newcommand{\gbr}{{{{g_{\rm br}}}}}
\newcommand{\gff}{{{{g_{\rm ff}}}}}
\newcommand{\xe}{{{{x_{\rm e}}}}}
\newcommand{\alphafs}{{{{\alpha_{\rm fs}}}}}
\newcommand{\YHe}{{{{Y_{\rm He}}}}}
\newcommand{\SE}{{{\dot{{\mathcal{E}}}}}}
\newcommand{\SQ}{{{{{\mathcal{E}}}}}}
\newcommand{\SN}{{\dot{\mathcal{N}}}}
\newcommand{\Sn}{{{\mathcal{N}}}}
\newcommand{\muc}{{{{\mu_{\rm c}}}}}
\newcommand{\xc}{{{{x_{\rm c}}}}}
\newcommand{\xH}{{{{x_{\rm H}}}}}
\newcommand{\mT}{{{{\mathcal{T}}}}}
\newcommand{\Ob}{{{{\Omega_{\rm b}}}}}
\newcommand{\Or}{{{{\Omega_{\rm r}}}}}
\newcommand{\Odm}{{{{\Omega_{\rm dm}}}}}
\newcommand{\mdm}{{{{m_{\rm WIMP}}}}}
\newcommand{\Acmb}{{{{A_{\rm CMB}}}}}
\newcommand{\Ayco}{{{{A_{\rm y/CO}}}}}
\newcommand{\Ad}{{{{A_{\rm dust}}}}}
\newcommand{\Td}{{{{T_{\rm dust}}}}}
\newcommand{\betad}{{{{\beta_{\rm dust}}}}}
\newcommand{\fyco}{{{{f_{\nu}^{\rm y/CO}}}}}
\newcommand{\nudo}{{{{\nu_{0}^{\rm dust}}}}}
\newcommand{\fnud}{{{{f_{\nu}^{\rm dust}}}}}

\title{Limits on the fluctuating part of $y$-type distortion monopole from Planck and SPT results}

\author[a]{Rishi Khatri,}
\author[a,b,c]{Rashid Sunyaev}

\affiliation[a]{ Max Planck Institut f\"{u}r Astrophysik\\, Karl-Schwarzschild-Str. 1
  85741, Garching, Germany }
\affiliation[b]{Space Research Institute, Russian Academy of Sciences, Profsoyuznaya
 84/32, 117997 Moscow, Russia}
\affiliation[c]{Institute for Advanced Study, Einstein Drive, Princeton, New Jersey 08540, USA}
\date{\today}
\emailAdd{khatri@mpa-garching.mpg.de}
\abstract
{We use  the published
   Planck and SPT cluster catalogs \cite{planckclusters2015,spt2015} and recently published  $y$-distortion
   maps \cite{k2015}  to put strong
  observational
  limits on the contribution of the fluctuating part of the $y$-type
  distortions to the $y$-distortion monopole. Our bounds are $5.4\times
  10^{-8} < \langle
  y\rangle < 2.2\times 10^{-6}$. Our upper bound is a factor of 6.8
  stronger than the currently best upper $95\%$ confidence limit from COBE-FIRAS of $\langle
  y\rangle <15\times 10^{-6}$.  In the standard cosmology, large
  scale structure is the only source of such distortions and our limits
  therefore constrain the baryonic physics involved in the formation of the
  large scale structure. Our lower limit, from the detected clusters in the
  Planck and SPT catalogs, also implies that a Pixie-like
  experiment  should
  detect the $y$-distortion monopole at $>27$-$\sigma$.  The biggest sources of uncertainty in our
  upper limit are the monopole offsets between different HFI channel maps
  that we estimate to be $<10^{-6}$.}

\maketitle
\flushbottom
\section{Introduction}
The Planck experiment \cite{planckmission} is the first full sky experiment after COBE (Cosmic
Background Explorer) \cite{cobe,cobedmr}  covering the entire
CMB spectrum  (the blackbody peak, the Rayleigh-Jeans and the Wien
regions) with high sensitivity. The wide frequency coverage combined with the
unprecedented sensitivity of Planck allows us to detect and separate the
$y$-type distortion \cite{zs1969} over the whole sky and Planck
collaboration as well as other groups have already created the maps of $y$-type
distortion from Planck data \cite{planckymap,hs2014}. The $y$-type
distortion spectrum is given by \cite{zs1969}
\begin{align}
\Delta I_{\nu}=\frac{2 h\nu^3}{c^2}\frac{x
  e^x}{(e^x-1)^2}\left[x\left(\frac{e^x+1}{e^x-1}\right)-4\right],
\end{align}
where $x=h\nu/(\kB\TCMB)$, $h$ is the Planck constant, $\kB$ is the
Boltzmann constant, $\TCMB=2.725$ is the CMB temperature and $c$ is the
speed of light. {The $y$-type distortion has negative intensity, compared to
the background CMB, at $\nu<217{\rm GHz}$ and positive intensity at
$\nu>217{\rm GHz}$. The Planck experiment has channels covering both regions
as well as at the null (217 GHz). {The South Pole Telescope (SPT) and the
Atacama Cosmology Telescope (ACT) cover
only the negative region and the null at present but at much higher angular resolution
of $\sim 1'$ \cite{spt2015,act2014}. The ACT experiment has also observed at
277 GHz and results including this channel should become available in the
near future \cite{act20142}.
}

Planck and other CMB
anisotropy experiments measure the changes in the brightness  as
they scan the sky and are insensitive to the  invariant part of
the signal. The $y$-type
distortion present in the Planck maps is therefore  expected in the standard
cosmology to be solely from the Compton scattering of blackbody CMB photons
by the hot electrons present in the clusters and groups of galaxies and hot filaments \cite{sz1972},
the result of cosmological structure formation. Since the electrons are
hotter than the CMB, Compton scattering results in transfer of energy from
electrons to photons, up-scattering the photons from the Rayleigh-Jeans part
to the Wien region of the spectrum. The result is a decrease in the
intensity of the CMB in the Rayleigh-Jeans part and an increase in the Wien
region of the CMB spectrum resulting in the characteristic spectrum of the
$y$-type distortion.\footnote{A detailed review of the various aspects of 
Compton scattering can be found in \cite{pss1983}.} 
The $y$-type distortion has been used by the SPT (South Pole Telescope),
ACT (Atacama Cosmology Telescope) and Planck to discover hundreds of new
clusters in addition to studying the physics of known clusters discovered
optically or through x-rays
\cite{sptcluster0,sptcluster1,actcluster2,elgordo,actcluster3,sptcluster3,planckclusters,planck2015}.

In the future we expect to have  experiments such as Pixie (Primordial
Inflation Explorer \cite{pixie}) or
PRISM ( Polarized Radiation Imaging and Spectroscopy Mission \cite{prism})
sensitive to the absolute brightness of the sky and not just the changes in
brightness as they scan the sky.  These experiments are expected  to be 3-4 orders of magnitude more
sensitive then COBE-FIRAS (Far Infrared Absolute Spectrophotometer)
\cite{fm2002,m2007}. These experiments would in particular measure the average
$y$-type distortion or the monopole on the sky improving by many orders of magnitude the
current COBE-FIRAS upper limit of $y<1.5\times 10^{-5}$ \cite{cobe}. 

{However since we already have multi-frequency experiments such as Planck and SPT which are
more sensitive than COBE-FIRAS and DMR (Differential Microwave Radiometer),} it is an
interesting question to ask if we can already put better limits on the
$y$-type distortion monopole? The answer turns
out to be yes and we use Planck and SPT data to put both upper and lower
limits on the $y$-distortion monopole. This is possible because the
$y$-type distortion from the large scale structure is a strictly positive very
inhomogeneous signal with a non-varying part that is very small compared to
the part which varies on the sky. Therefore the contribution to the
monopole or the average $y$-type distortion from the invariant part, to which the
experiments such as Planck and SPT are insensitive, is expected to be
negligible compared to the contribution from the fluctuating part of the
$y$-type distortions, at least in standard cosmological model without any new
physics.  In the standard cosmological model, there are two main contributors to the
invariant part of the $y$-type distortion. 

The first is the dissipation of
sound waves before and during recombination known as Silk damping \cite{silk,Peebles1970,kaiser,dzs1978} due to
photon diffusion and free streaming and results in an exponentially decaying
anisotropy spectrum in the CMB. This decrease in the power in sound waves
on small scales during the epoch of recombination has already been measured by SPT, ACT and Planck
\cite{sptdamping,actdamping,planck}. This energy which disappears from the
CMB anisotropies appears in the CMB monopole as spectral distortions \cite{sz1970,cks2012,ksc2012b}. At
$z\lesssim 10^4$ the distortion is a $y$-type distortion while at higher redshifts there is thermalization of the
distortions towards the equilibrium Bose-Einstein spectrum resulting in the
intermediate and $\mu$-type distortions
\cite{sz1970,is1975b,is1975,hs1993,ss1983,bdd1991,pb2009,cs2011,ks2012b,ks2013b}.
Since the Universe in the standard model is statistically homogeneous and
Gaussian, the sound waves on
small scales have the same amount of power everywhere. Therefore we expect
that the same amplitude of distortions would be created everywhere, i.e. we
would have an invariant contribution to the $y$-distortion monopole. There
will be a
small 
difference in different parts of the Universe because of the cosmic variance
which we can neglect. We should mention that in the presence of
non-Gaussianity we can have a spatially varying distortion since
non-Gaussianity can cause the small scale power to vary from place to place
\cite{pajer2012,ganc2012,edck2015}. The invariant $y$-type distortions
coming from the Silk damping are expected to be $y^{\rm Silk ~damping}\sim 4\times 10^{-9}$ \cite{cks2012} in
the standard $\Lambda$CDM cosmological model.  {Any non-standard energy release during the recombination, which may also
create $y$-type distortions, would also delay recombination and therefore
change the CMB anisotropy power spectrum. Any energy release between
recombination and reionization is therefore already tightly constrained by the CMB anisotropies
\cite{ck2004,pf2005,spf2009,hch2011,planck2015,cs2011,ks2012} and the resulting $y$-type distortion is constrained to be sub-dominant compared to the Silk damping signal.}

The second homogeneous contribution is expected to come from
{reionization at $z\gtrsim 6$.} We expect the temperature of gas during the reionization to
be $\sim 10^{4}{\rm K}$ \cite{bl2001} and with optical depth through reionization of
$\tau \sim  0.07$ \cite{planck2015}, we expect $y^{\rm
  Reionization}\approx\tau \Te/\me \sim 10^{-7}$ \cite{pixie}. 

{The fluctuating part on the other hand is expected to be dominated by the
contributions from clusters \cite[page 15]{sz1970c} and groups of galaxies and warm hot
intergalactic medium (WHIM) \cite{sz1972b,co1999}. The contribution from
the peaks in the large scale structure and
the effect of baryonic physics to the $y$-type distortion
fluctuations has been studied by many groups
using numerical simulations as well as analytically
\cite{Refregier2000,springel1,springel2,ns2001,white2002,schaefer2006,sn2010,bbp2010,bbps2012,Munshi2013,ds2013,dsk2015}.
The predictions for the average distortion from this fluctuating component range from
$y=10^{-6}-3\times 10^{-6}$
\cite{Refregier2000,springel1,springel2,ns2001,white2002,schaefer2006,bbp2010,Munshi2013,dsk2015}
with majority of simulations leaning towards
$\gtrsim 2\times 10^{-6}$. This average distortion is sensitive to baryonic
physics, in particular to any mechanism that can inject energy into the
intergalactic medium. These  predictions are the ones that the Planck
experiment is sensitive to and are possible to test with the current data.}

In addition to the anisotropic contribution to the $y$-distortion from the
large scale structure, we also have contribution from the hot gas in our
vicinity. Our Solar system is embedded in a local cavity or bubble of $s\approx
40-130{\rm pc}$ size filled
with  hot ionized gas
 with temperatures $\approx 10^6 ~{\rm K}$ and
pressure of $P/\kB\approx 1.5\times 10^4~{\rm cm^{-3}K}$ \cite{sef1998}. This implies a local $y$-type
distortion from our vicinity of $y= \sigT P d/(\me c^2)\approx 5\times 10^{-10}$, where
$\sigT$ is the Thomson scattering cross section and $\me$ is the electron mass.
Our galaxy is also surrounded by a hot halo of ionized gas belonging to  the warm hot
circumgalactic medium  and the local group medium
with temperature $\approx 2\times 10^{6}{\rm K}$ and electron density
$\Ne\approx 2\times 10^{-4}~{\rm cm}^{-3}$  extending to
$\sim 72~{\rm kpc}$ \cite{gmkn2012} giving $y\sim 5\times 10^{-9}$. These small local
contributions may be detectable by future experiments and could be separated from
the cosmological $y$-distortion owing to their characteristic anisotropic
signature on the sky.

\section{The  full sky map of $y$-type distortion and masks}
{Planck collaboration maps  \cite{planckymap} of the $y$-type distortion created using
the internal linear combination methods are now publicly 
available and use two different internal linear combination (ILC)
algorithms MILCA (Modified ILC Algorithm) \cite{milca} and NILC (Needlet
ILC) \cite{nilc}.}
We will use the maps of $y$-type distortion constructed using the Planck
HFI channels in \cite{k2015} using a different method. These maps are constructed using the
linearized iterative least-squares (LIL) parametric model fitting scheme
proposed in \cite{k2014}. We reproduce the $y$-map from \cite{k2015} in
Fig. \ref{Fig:ymap}. The well known clusters can easily be spotted on the
map. Approximately $14\%$ of the sky most contaminated by the dust and CO
emission has been masked in the map. 
\begin{figure*}
\resizebox{\hsize}{!}{\includegraphics{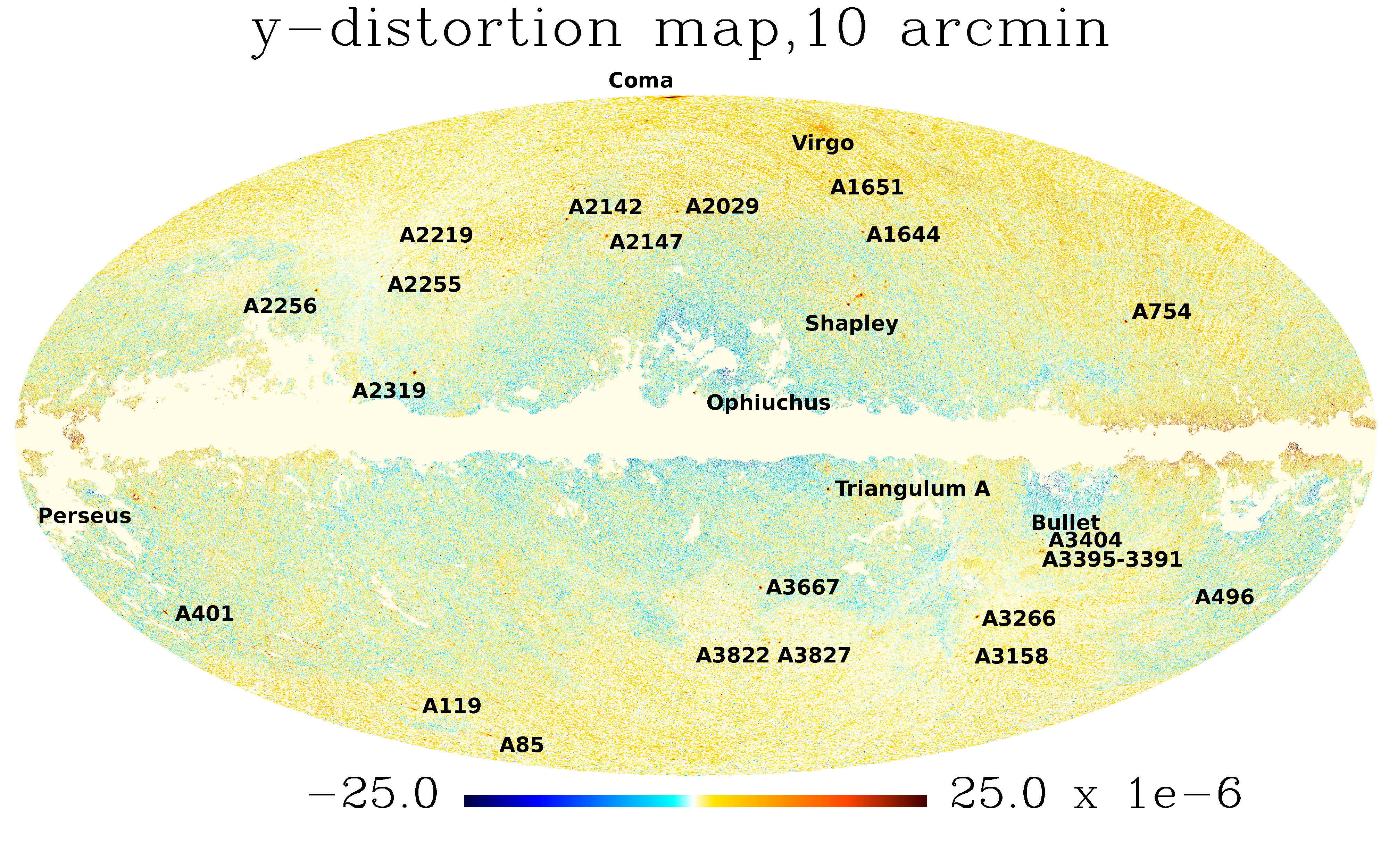}}
\caption{\label{Fig:ymap}Map of $y$-type distortion at $10'$ resolution
  constructed from the lowest four Planck HFI channels in \cite{k2015}.}
\end{figure*}

An advantage of the parametric model fitting is
that it gives a quantitative estimate of how good the model fits the
data in the form of $\chi^2$. This information was used in  \cite{k2015} to construct a mask which
masks the regions on the sky worst affected by the carbon-monoxide (CO)
emission, the main contaminant for the $y$-distortion signal. 
 
The Planck
catalog was also revisited in \cite{k2015} and clusters and cluster
candidates  where the $y$-type
distortion signal should be free of CO
contamination were identified. We will use this CO mask and the clean
sample of clusters to put reliable upper and lower limits on the average
$y$-type distortion. We will try to be conservative everywhere, so that
small amount of remaining contamination would always expand our limits. We
consider the
limits derived in this ultra-conservative way to be the hard limits and the
probability that these limits are violated should be vanishing. In
calculating the upper limit we will directly average the $y$-type distortion in
the pixels in $y$-map in real space. We refer the reader to \cite{k2014} for details
of our component separation method and to \cite{k2015} for details of the
construction of the $y$-type distortion map, CO mask and identification of
clean clusters. An important feature of our algorithm is nested
  model selection. We fit a 3 parameter CMB+dust model as well as a
  CMB+dust+y model to every pixel. The difference in $\chi^2$ (or
  $-2\ln({\rm Likelihood})$) for the 3 parameter and 4 parameter model has
  a $\chi^2$ distribution with one degree of freedom. We set a threshold in
  the $\chi^2$ difference for the two fits, $\Delta \chi^2$, for the
  acceptance of the extra $y$ parameter. If the improvement in $\chi^2$ by
  adding the extra $y$-type component to the sky emission model is larger
  than the threshold $\Delta \chi^2$, we accept that the $y$ component is
  present and take the best fit amplitude. If the improvement is smaller
  than the threshold we reject the $y$-type component and set the $y$
  amplitude in that pixel to zero. A $\Delta \chi^2=0$ threshold would
  imply no model selection and accepting the $y$-type component everywhere
  while thresholds of $1.6,2.7,3.8$ imply that there is $20,10,5\%$
  probability respectively that we except the 4 parameter model with $y$ component  when
  the fourth $y$-component is in fact absent i.e. these are the probabilities for
 the  false detection of $y$-type signal. Higher thresholds would therefore
result in cleaner $y$-maps with the price that some of the below noise
$y$-signal would also be lost. The map shown in Fig. \ref{Fig:ymap}
corresponds to $\Delta \chi^2=3.8$.

In addition to the Galactic CO emission, the CO emission lines from external galaxies form a diffuse background. This background, integrated over all redshifts, is expected to
  contribute $\sim 1~{\rm \mu K}$ to Planck frequency channels at 100 GHz
  and above with an almost flat spectrum \cite{rhs2008}. The variation with
  frequency in the redshift integrated spectrum,
  which is relevant for contamination to the $y$-distortion, should be even
  smaller, of order $0.1 \sim \mu{\rm K}$ and therefore the contamination
  to the average $y$-type distortion of $< 10^{-7}$.\footnote{For Planck 100 GHz
  channel the conversion
  from $y$ to brightness temperature $K_{\rm CMB}$ units is $T_{100 ~{\rm GHz}}=-4.031 y$ \cite{planckhfi}.} This is a
  factor of 20 smaller than the upper limit we will get and we can neglect
  it. Note that the
  diffuse extragalactic background can give either positive or negative
  contamination to the $y$-distortion depending on its exact spectral shape. The
  galactic CO emission on the other hand, with the canonical line ratios
  that we have assumed, 
  always gives a positive contamination and will bias our upper limits
  towards higher values. 
 The strongest contaminant for us is the
  Galactic CO and dust emission and we will focus on these from now on.

\subsection{Comparison with Planck maps}
{We show in Fig. \ref{Fig:lilpdf} the probability distribution
  function (PDF, $P(y)$) of our map calculated with LIL algorithm
  for $\Delta \chi^2=0.0$ and 
  compare it with the MILCA and NILC maps in Fig. \ref{Fig:pdf} for
  $51\%$ sky fraction (masks are defined in the next section). 
Also shown
  are the $P(y)$ when all clusters and cluster candidates in the second Planck cluster catalog are
  masked. We see in Fig. \ref{Fig:lilpdf} the skewness in the PDF compared
  to the symmetric and Gaussian noise PDF, as predicted by \cite{rs2003},
  from the non-Gaussian $y$-distortion signal not only in the far tail
  coming from massive clusters, but also at small $y$ near the center,
  coming from unresolved clusters and groups of galaxies.
All maps in Fig. \ref{Fig:pdf}
  agree in the positive tails of the distribution which are dominated by
  the detected clusters. The LIL map  has larger noise (calculated from the half
  ring half difference maps) compared to the MILCA and
  NILC maps. The width of the MILCA and NILC distributions are however
  significantly wider than their noise distributions indicating that their
  maps are probably dominated by contamination while our maps are dominated
  by noise.  The most significant difference in our maps and Planck maps is
  the absence of positive skewness at small values of $y$ in the MILCA and
  NILC maps. We expect this skewness 
  if there was $y$-signal present in the maps below the pixel noise level in the maps. This
  is probably because the Planck ILC based algorithms explicitly remove the monopole,
  the very signal we are interested in. We therefore cannot use MILCA and
  NILC maps to estimate the average $y$-distortion signal and will only
  present results from LIL maps. The negative tails, representing galactic
   and radio source contamination, are slightly larger in case of
   LIL. However the contribution of these tails to our average $y$-signal
   is negligible since their probability is $< 10^{-3}$ compared to
   the peak. The average $y$-signal corresponds to the integration over the
   PDF and gets most of the contribution from near the peak of the PDF at
   $|y|\sim {\rm few}\times 10^{-6}$. We will also quote ultra-conservative values, integrating over only the
   positive half of the PDF explicitly ignoring the negative contamination tail.
}
\begin{figure}
\resizebox{\hsize}{!}{\includegraphics{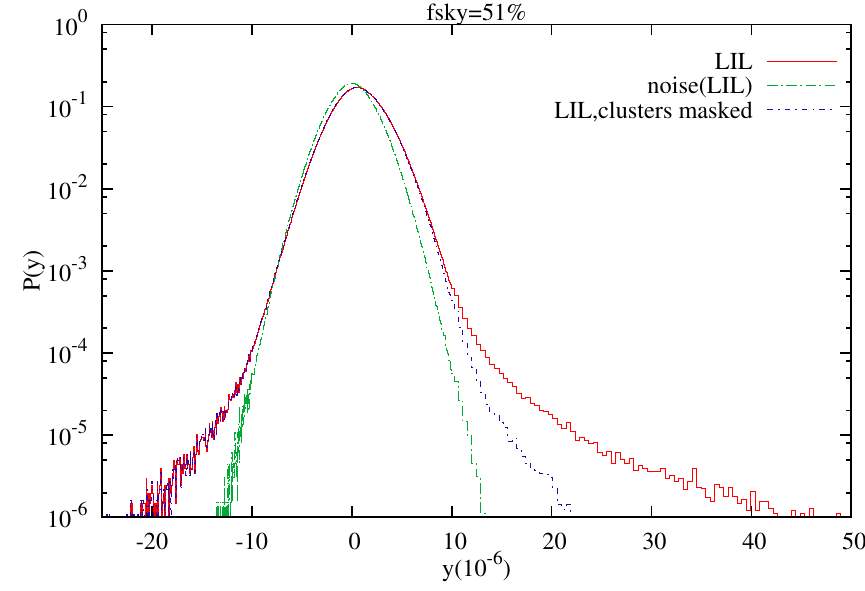}}
\caption{\label{Fig:lilpdf}The PDF of our $y$-distortion map calculated
  with LIL algorithm for the
  same $51\%$
of the sky. }
\end{figure}
\begin{figure}
\resizebox{\hsize}{!}{\includegraphics{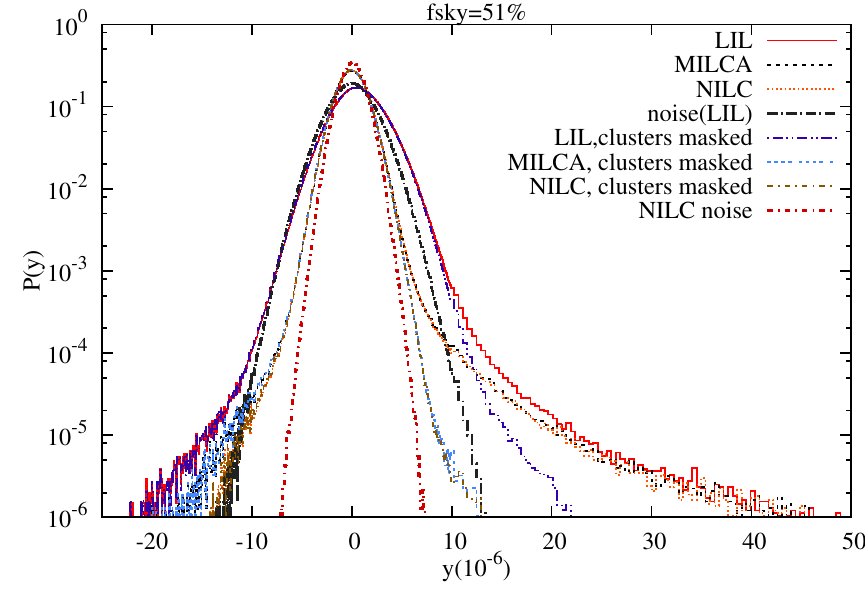}}
\caption{\label{Fig:pdf}The PDF of LIL,NILC, and MILCA $y$-maps for the
  same $51\%$
of the sky. Note that the difference from the PDF published in
\cite{planckymap} is because they applied additional filtering to their
maps before calculating the PDFs while we plot the unfiltered PDFs.}
\end{figure}

\section{Upper limits on the average $y$-distortion}
We will use the $y$-map and CO masks to estimate the
average $y$-type distortion or the monopole. We have seen in the previous
sections that we should expect non-negligible contamination from dust and CO
emission in the $y$-map, especially for pixels where no cluster is
detected. Therefore even though it is trivial to calculate the
$y$-distortion from the $y$-map with statistical error that should be extremely
small 
owing to the averaging over the millions of pixels in the whole sky, the systematic error is much
harder to estimate. We will therefore take a conservative approach and
instead of giving the error bar at some number of $\sigma$, we will try to
put a hard upper bound so that the probability that the signal would exceed
our upper bound is vanishing. We will validate our results using the FFP6
simulations \cite{ffp62013}.

We  augment our minimal $86\%$ mask using the 545 GHz channel to
get successively cleaner and cleaner portions of the sky. Note that unlike
the non-zero low order multipoles, for the
monopole we do not need a large fraction of the sky and  we
should expect the monopole to be same in different small patches of the
sky for a statistically homogeneous and isotropic Universe. We also
analyze the maps generated with different nested model selection
thresholds of $\Delta \chi^2=0,1.6,2.7,3.8$ to verify how much signal is
lost by our model selection procedure, $\Delta \chi^2=0$ means no model
selection and we fit for the $y$-distortion component everywhere. Even
though we have taken care to mask strong radio sources which appear as
negative $y$ sources in the $y$-map, there will still be some sources below our
thresholds that would
remain in the map. There is of course positive foreground contamination
also, mostly from the residual CO emission, the positive and negative
contamination is not expected however to  completely cancel each
other. There are two ways to handle these pixels when we are interested in
the upper bounds.  A \emph{conservative} approach would be to set
all negative pixels to zero and then average over all the pixels. This  procedure would bias our
results towards higher values of $y$ as the positive noise fluctuations are
still added to the signal. An \emph{ultra-conservative} approach would be
to mask or  completely ignore the negative pixels and average only the
positive pixels completely avoiding the dilution in signal from the
negative pixels. In what follows we take the ultra-conservative
approach. The resulting  $y$-type monopole  for
different combinations of above mentioned analysis procedures is summarized
in Table \ref{Tbl:yupper}. We will label the average $y$ when keeping the
negative values as \emph{average estimate} and results when masking negative
pixels  as \emph{Ucons. estimate}. The conservative values of course lie
between these two extremes. 

\begin{table}
\begin{tabular}{|c|c|c|c|c|}
\hline
  $\Delta \chi^2$  &$0.0$ &$1.6$&$2.7$&$3.8$ \\
                 &average | Ucons.&average | Ucons.&average |
                 Ucons.&average | Ucons.\\
\hline
$81\%$ sky&0.49 | 2.32&0.53 | 0.95&0.56 | 0.94&0.57 | 0.94 \\
w/o clusters&0.44 | 2.28&0.50 | 0.91&0.52 | 0.91&0.53 | 0.91\\
\hline
$71\%$ sky&0.57 | 2.26&0.60 | 0.96&0.62 | 0.96&0.64 | 0.96\\
w/o clusters&0.52 | 2.22&0.56 | 0.93&0.59 | 0.93&0.60 | 0.92\\
\hline
$61\%$ sky&0.62 | 2.25&0.67 | 1.00&0.70 | 0.99&0.71 | 0.99\\
w/o clusters&0.57 | 2.21&0.64 | 0.96&0.67 | 0.96&0.68 | 0.95\\
\hline
$51\%$ sky&0.66 | 2.24&0.75 | 1.03&0.79 | 1.02&0.80 | 1.02\\
w/o clusters&0.62 | 2.20&0.72 | 1.00&0.75 | 0.99&0.77 | 0.99\\
\hline
$41\%$ sky&0.68 | 2.22&0.82 | 1.06&0.86 | 1.05&0.88 | 1.05\\
w/o clusters&0.63 | 2.18&0.79 | 1.03&0.83 | 1.02&0.85 | 1.01\\
\hline
$31\%$ sky&0.65 | 2.16&0.87 | 1.08&0.92 | 1.06&0.95 | 1.06\\
w/o clusters&0.60 | 2.12&0.83 | 1.04&0.88 | 1.03&0.91 | 1.02\\
\hline
\end{tabular}
\caption{\label{Tbl:yupper}Values of $y$-distortion monopole amplitude
  $\langle y\rangle \times 10^{6}$. Both the average values (average) and
  Ultra-conservative upper bounds (Ucons.) are presented.} 
\end{table}

There is a small systematic variation with the changing sky fraction and
$\Delta \chi^2$, which is expected, in addition to a jump when the model
selection is turned on. The fact that the changes from $\Delta \chi^2 =
1.6$ to $3.8$ are very small gives us confidence that our model selection
is not throwing away too much signal along with the contamination. In particular both
the
average and ultra conservative values converge towards $\langle
y\rangle=1.0\times 10^{-6}$ with the decreasing sky fraction and increasing
$\Delta \chi^2$ signaling the decrease in foreground contamination.
We therefore take $\langle
y\rangle=1.0\times 10^{-6}$ as our best estimate for the average $y$-type
distortion and take the $\Delta \chi^2=0.0,31\%$ sky fraction value of
$2.2\times 10^{-6}$ as hard upper bound. It is clear from the table that
this is a very conservative bound and we expect that the probability that
the average $y$-type distortion will exceed this value is vanishing. We
note that this bound  is a factor of 6.8 times stronger the COBE $95\%$ upper limit of
$15\times 10^{-6}$ \cite{cobe}. Our limits of course only apply to the
fluctuating contribution whereas COBE-FIRAS limits are more generally applicable.
We should also mention that for  $\Delta \chi^2=0.0,31\%$ sky fraction the
conservative limit (see above) is $1.3\times 10^{-6}$ close to our best estimate.

We also note that the contribution of clusters from the Planck catalog
outside our masks are consistent across the table at $\langle
y\rangle_{\rm clusters}\approx 4\times 10^{-8}$ and this
provides a strong lower limit to the $y$-type distortion. We will compare
this limit with one derived directly from the Planck and SPT cluster
catalogs in the next section.

\subsection{Effect of systematic offsets between channels, validation  with FFP6 simulation and with lower resolution maps}
The FFP6 simulations simulate the nominal Planck sky and therefore have
higher noise compared to the Planck full mission. Nevertheless we can use
the FFP6 simulations to answer the following question: Do the foregrounds
bias our results towards higher or lower values of $y$. We have repeated our
analysis on the FFP6 simulations which have an input average $y$ of
$\approx 3.9\times 10^{-7}$. Our output $y$-map
has $\langle y\rangle = 4.5\times 10^{-6}$ corresponding to the
ultra-conservative upper limit on $31\%$ sky and no model selection and
$\approx 2.8\times 10^{-6}$ corresponding to the $\Delta\chi^2=3.8$. This
values are considerably higher compared to our values on the real
sky and the true value. 

The main 
reason is that in the FFP6 simulations there are offsets present of order
$10-30~{\mu \rm K}$ between the maps which give an almost constant monopole $y$-signal
over the whole sky in our reconstructed $y$ map. This offset contribution,
which is positive $y$-monopole, is really what we are
measuring as the average $y$. To test this, we remove 
the relative offsets from
the FFP6 channel maps estimated from the CMB only channel maps. After
removing the rough offsets we get ultra conservative limits of $\langle y\rangle
= 1.64\times 10^{-6}$ and $1.0\times 10^{-7}$ for $\Delta\chi^2=0.0$ and
$3.8$ respectively. Thus with and without model selection values nicely
bracket the true value of $\langle y \rangle = 3.9\times 10^{-7}$.

For the real data the offsets have been
estimated and removed in the released maps \cite{hfical2015}. We also removed the
CIB monopole that is added to the released Planck maps together with the best
estimates of the residual dipoles \cite{comp2015}. If we do not remove the
CIB monopoles, we get much higher values of $\langle y\rangle =4.1\times
10^{-6}$ and $2.6\times 10^{-6}$ for  $\Delta\chi^2=0.0$ and
$3.8$ respectively. The maximum estimated uncertainty in
the 100 GHz, 143 GHz and 217 GHz channels for the residual offsets is $3-5~\mu{\rm
  K}$ and it is $23~\mu{\rm
  K}$ for the 353 GHz channel \cite{hfical} and these are the biggest source of
uncertainties for our upper limit. {We re-run our calculations
  after adding offsets corresponding to these uncertainties to one channel
  at a time. The results for $\Delta \chi^2=0.0$ and sky fraction $f_{\rm sky}=31\%$ are tabulated in Table \ref{Tbl:sys}. We therefore
  expect the maximum  error in our
$y$-estimate  to be of order
$\sim  10^{-6}$ assuming that the offsets do
not completely mimic the $y$-distortion. If the offset errors are random
they will partially cancel each other and we should expect the actual error
to be smaller.} 
\begin{table}
\begin{tabular}{|c|c|c|c|c|}
\hline
&100 GHz&143 GHz&217 GHz&353 GHz\\
\hline
offset $(\mu K)$&-3&-3&+5&+23\\
\hline
average y ($\times 10^{-6}$)&0.66&1.92&3.0&-0.39\\
Ucons. y ($\times 10^{-6}$)&2.2&2.76&3.46&1.83\\
\hline
\end{tabular}
\caption{\label{Tbl:sys}Effect of introducing systematic offsets in one
  channel at a time. The signs of offsets are chosen to be in the same direction as the
  $y$-distortion signal. Note that for the 353 GHz channel, the offset is
  getting absorbed in the other components and $y$-distortion signal
  compensates by going in the direction opposite than expected.} 
\end{table}

The FFP6 simulations do not contain diffuse emission from the WHIM \cite{co1999}
that we expect in the real sky. If such diffuse emission is present, it
should be possible to place better limits on the diffuse emission by
rebeaming the maps to a bigger beam and redoing the analysis. We do this
for the real full mission maps at $30'$ and $60'$ resolution. Note that
once we are at such low resolutions it no longer makes sense to do model
selection since the signal will be spread out over large areas. We therefore
turn-off the model selection (i.e. set $\Delta \chi^2=0$) and redo the
analysis as in the previous section. For $30'$ resolution we get a
conservative $\langle y\rangle=4\times 10^{-7}$ and an ultra-conservative
value of $7.4\times 10^{-7}$. The corresponding values for the $60'$
resolution are $2.6\times 10^{-7}$ and $5.4\times 10^{-7}$. There is
therefore a trend towards lower values with increase beam
size. {For these large beams however there is not sufficient
  control over the foregrounds which also spread out.  We therefore use the
  smallest average $\Delta \chi^2=0,f_{\rm sky}=31\%$ and largest average
value for  $\Delta \chi^2=3.8,f_{\rm sky}=31\%$ from Table \ref{Tbl:yupper} to give our best estimate
for average $y$ to be $7\times 10^{-7}\le \langle y\rangle \le 10^{-6}$.} This discussion also shows the difficulty of actually measuring
the average $y$-type distortion in the Planck maps and this is the reason
we opt instead to put conservative upper and lower bounds.

\section{Lower limits  on the average $y$-distortion from the Planck and
  SPT cluster catalogs}\label{sec:llimit}
There are already more than 1000 clusters detected by Planck and SPT and
available as catalogues. The average $y$ distortion of course includes
contribution from these clusters and therefore we can use these catalogs to
provide a hard lower bound on the average $y$ distortion. We use both the
Planck and SPT catalogs since they are complimentary. Planck is very good
at detecting large nearby clusters while SPT is more efficient in detecting
the distant clusters. We use the second Planck catalog
\cite{planckclusters2015} covering $83.6\%$ of the sky and the SPT catalog
covering 2500 ${\rm deg}^2$ \cite{spt2015}. 

{The Planck catalog gives the  values of $y$ integrated over the cluster $Y_{500}$
 within the radius $R_{500}$, i.e the radius inside which the average
 density is 500 times the critical density at the redshift of the
 cluster, estimated from  measurements of $Y_{5R500}$ assuming a cluster model.} The SPT catalog on the other hand provide the core radius
 ($\theta_c$) of the
 beta profile fit to the cluster \cite{cf1976} along with the integrated
 $y$ signal inside the $\theta_{\rm max}=0.75'$ radius for each cluster,
\begin{align}
Y({\theta_{\rm max}})&=2\pi \int_0^{\theta_{\rm
    max}}{\frac{y_0\theta\id\theta}{1+\theta^2/\theta_c^2}}\nonumber\\
&=\pi y_0 \theta_c^2\log\left(1+\frac{\theta_{\rm max}^2}{\theta_c^2}\right)
\end{align}
The beta profile has been previously found to be a good fit to the SPT
clusters out to virial radius of $\theta_{500}\approx 5 \theta_c$
\cite{spt2010}. We therefore integrate the $y$ signal out to
$5\theta_c$ for all SPT clusters. The average contribution of Planck and
SPT clusters is given in Table \ref{Tbl:llimit}. 

\begin{table}
\begin{tabular}{|c|c|c|c|c|c|c|}
\hline
  sample  &\multicolumn{4}{|c|}{Planck} &\multicolumn{2}{|c|}{SPT}\\
\hline
          &{all}&{ S/N$>6$}& {clean} &{clean(S/N$>6$)}&$3\theta_c$ &$5\theta_c$\\ 
\hline
$z\le 0.3$&0.038 &0.029&0.033&0.025&0.0073&0.010 \\
$z>0.3$&0.019&0.0074&0.0099&0.0038&0.021&0.030\\
\hline
\end{tabular}
\caption{\label{Tbl:llimit}Average values of $y$-distortion monopole amplitude
  $\langle y\rangle \times 10^{6}$ calculated from the Planck and SPT
  cluster catalogs \cite{planckclusters2015,spt2015}. We divide the samples
  into low redshift and high redshift samples at $z=0.3$. The Planck and
  SPT clusters give comparable contributions to the average $y$-distortion. } 
\end{table}

For the clean Planck sample, where we remove the sources identified by our
algorithm as molecular clouds \cite{k2015}, we get a total value of $4.3\times 10^{-8}$
which agrees beautifully with the value we estimated directly from our
maps for the Planck clusters in the previous section. This shows indirectly
the agreement between our $y$-maps and those created by the Planck
collaboration. From Fig. 6 in \cite{spt2015} we see that most of the Planck
detected clusters lie at $z<0.3$ whereas the majority of the SPT clusters
are at $z>0.3$. We therefore use this redshift to divide both the SPT and the
Planck clusters into the low redshift and high redshift samples. For the
low redshift contribution to the average $y$-distortion we take the Planck
low redshift clean sample value of $\langle y\rangle_{z\le 0.3}=3.3\times
10^{-8}$. In calculating this value we have not included the clusters for
which redshift is not known. Those clusters are included in the $z>0.3$ values.

For the SPT samples, we show values calculated by extrapolating the quoted
parameters in the SPT cluster to $3\theta_c$ and $5\theta_c$. For the high
redshift sample, there is a difference of $30\%$ in these two values. For
our lower limit we decide to be conservative and choose the smaller value
giving  $\langle y\rangle_{z> 0.3}=2.1\times
10^{-8}$. Combining the Planck and SPT therefore we get the final lower
bound on the average $y$ distortion of $\langle y\rangle >5.4\times
10^{-8}$. In the published proposal of the Pixie experiment \cite{pixie},
the target is given as $5$-$\sigma$ detection of $\langle y\rangle =
10^{-8}$. Our calculations therefore guarantee that Pixie would detect an average
$y$-type distortion at least $27$-$\sigma$.
 
\section{Conclusions}
We have used our recently constructed  maps of $y$-type distortion from the
Planck HFI channels to arrive at strong upper bounds on the average
$y$-distortion or the monopole. The FFP6 simulations indicate that our
value is dominated by the foregrounds contamination which drives the value
higher and this bound should be considered as hard upper bound. This bound
is $\langle y\rangle <2.2\times 10^{-6}$ and is a factor of 6.8 stronger
than the COBE-FIRAS $95\%$ upper limit of $15\times 10^{-6}$. There is however a
caveat. Our limit only applies to the contribution to the monopole from the
fluctuating component of the $y$-type distortion, since this is the signal
to which Planck is sensitive. The COBE-FIRAS measurements were on the other hand
absolute measurements and also constrain the non-fluctuating or the
invariant background
component. The two main sources of uncertainty in our $\langle y \rangle$
are the residual foregrounds, especially the galactic CO emission, and
offsets between the Planck channel maps since Planck does not measure the
absolute brightness of the sky. We estimate the error in our upper limit from the systematic
offsets to be $<10^{-6}$.

We have also derived a strong lower bound on the average $y$-type
distortion  using the Planck and SPT cluster catalogues. This bound is
$\langle y\rangle >5.4\times 10^{-8}$ and should also be considered a hard
lower bound.

Our analysis predicts a very optimistic picture for the proposed experiment
Pixie which should detect the global $y$-type distortion monopole at
$\gtrsim 27$-$\sigma$. We of course expect the actual distortions to be
much higher. With the future cluster surveys similar to current SPT, Planck
and ACT, we expect to measure and remove most of the contributions from the
low redshift large scale structure from the Pixie measurements and reach the
invariant (non-fluctuating) $y$-distortion created during the reionization.

\acknowledgments
This paper used observations obtained with Planck
(\url{http://www.esa.int/Planck}), an ESA science mission with instruments and
contributions directly funded by ESA Member States, NASA, and Canada. We
 also acknowledge  use of the HEALPix software \cite{healpix}
 (\url{http://healpix.sourceforge.net}) and FFP6 simulations generated
 using  the Planck
 sky model \cite{ffp62013}
 \url{http://wiki.cosmos.esa.int/planckpla/index.php/Simulation_data}.
 This research has made use of "Aladin sky atlas" developed at CDS,
 Strasbourg Observatory, France \cite{aladin}. This research has also made use of the SIMBAD database, operated at CDS, Strasbourg, France.
We would also like to thank
 Jacques Delabrouille and Julian Borrill for clarifying the presence of
 monopole offsets in the Planck FFP simulations.
RS acknowledges partial support by grant No. 14-22-00271 from the Russian Scientific Foundation.

\bibliographystyle{unsrtads}
\bibliography{ymonopole}
\end{document}